\documentclass[12pt, a4paper]{article}

\usepackage[
  margin=0.75in,
  headsep=10pt, 
]{geometry}

\usepackage{graphicx}
\usepackage{soul}
\usepackage{graphicx}
\usepackage{float}
\usepackage{epstopdf, epsfig}
\usepackage{amsmath}
\usepackage[table]{xcolor}
\usepackage{subcaption}
\usepackage{soul}
\usepackage{makecell}
\usepackage{multirow}
\usepackage{breqn}
\usepackage{amssymb} 
\usepackage{booktabs}
\usepackage{dcolumn}
\usepackage{bm}
\usepackage[utf8]{inputenc}
\usepackage[T1]{fontenc}
\usepackage{mathptmx}
\usepackage{etoolbox}
\usepackage{hyperref}
\usepackage{subcaption}
\usepackage{ragged2e}
\usepackage[sort&compress]{natbib} 

\newcommand{\edt}[1]{{\color{black}#1}} 

\newcommand{\soptitle}{Vortex Dynamics from Burst-and-Coast Motion of Anguilliform and Carangiform Swimmers}
\usepackage{xcolor} 

\begin{document}


\begin{center}
\Large \bf{\soptitle}
\vspace{0.1in}
\end{center}

\begin{center}
{Zahra Maleksabet$^1$, Maham Kamran$^1$, Ali Tarokh$^2$, and Muhammad Saif Ullah Khalid$^{1\ast}$}\\
\vspace{0.1in}
\end{center}
\begin{center}
$^1$Nature-Inspired Engineering Research Lab (NIERL), Department of Mechanical and Mechatronics Engineering, Lakehead University, Thunder Bay, ON P7B 5E1, Canada\\
$^2$Department of Mechanical and Mechatronics Engineering, Lakehead University, Thunder Bay, ON P7B 5E1, Canada\\
\vspace{0.05in}
$^\ast${Corresponding Author, Email: mkhalid7@lakeheadu.ca}
\end{center} 

\begin{abstract}

Fish perform various propulsive maneuvers while swimming by generating traveling waves along their bodies and producing thrust through tail strokes. Anguilliform swimmers spread motion along the body, while carangiform \edt{swimmers' motion is more prominent near their tails}. Many species also switch between continuous undulation and intermittent swimming, such as burst-and-coast maneuver, which can save energy but can also change the wake structure and \edt{hydrodynamic} forces. {Our current study aims at explaining} how duty cycle ($\mbox{DC}$), undulatory gaits, and Strouhal number ($\mbox{St}$), shape the near-body vortices, overall wakes, and the hydrodynamic forces. We \edt{carry out} three-dimensional simulations at $Re = 3000$ for \edt{flows around} an eel (anguilliform) and \edt{a} Jack Fish (carangiform) for $\mbox{DC} = 0.2-1.0$ and $\mbox{St} = 0.30$ and $0.40$. \edt{Our results reveal that the burst-and-coast motion for both swimmer produce bow-shaped wakes, the two rows of which on the sides approach each other to form a more coherent wake as $\mbox{DC}$ is increased to $1.0$ that corresponds to the wake of continuously undulating swimmers. It is also found that the intermittent motion at a higher Strouhal number produces more drag, contrary to the continuous undulatory kinematics. We further investigate this behavior by quantifying the strengths of vortices produced around the two swimmers and their instantaneous kinematic metrics. A detailed analysis for the role of different body sections in the production of unsteady streamwise forces is also presented. These insights provide important connections between the swimmers' physiologies, their kinematics, and the governing vortex dynamics to attain certain hydrodynamic metrics for designing next-generation autonomous bio-inspired underwater robots.}


\end{abstract}

\section{Introduction}
\label{sec:Intro}
\edt{For the past several decades,} aquatic locomotion \edt{is} a subject of intensive \edt{scientific investigations,} because it couples biology and fluid \edt{dynamics} for a single, richly unsteady phenomenon. Fish achieve efficient propulsion through body–flow interactions that redistribute momentum via coherent vortices \edt{formed} along the body and \edt{shed} from the caudal fin \edt{in the wake}. \edt{The principles and lessons learnt from hydrodynamics of fish,} inform the design of bio‑inspired underwater vehicles and autonomous robots that must balance endurance, maneuverability, and controllability in complex environments. Intermittent gaits are especially attractive from an energy‑management perspective, and a deeper understanding of how near‑body vortices organize thrust, drag, and side forces can translate directly into actuation strategies for fast and efficient propulsion in engineered systems.

Among undulatory strategies, anguilliform and carangiform swimming \edt{modes} represent two canonical gait families distinguished by both morphology and kinematics. Anguilliform swimmers like eels undergo large‑amplitude undulations distributed almost uniformly along the full body-length, generating thrust and creating intricate, spatially and temporally evolving flow features along the body and in the wakes \cite{miller2004vortices}. Simulations and experiments \edt{showed} that this distributed actuation \edt{promoted} the formation, convection, and interaction of body‑attached vortices \edt{to modulate} pressure over extended surface regions, yielding unsteady but sustained propulsive output \cite{borazjani2009numerical, khalid2021larger}. \edt{Contrarily}, carangiform swimmers like Jack fish concentrate most of the undulatory motion in the posterior \edt{part} of the body. The anterior body remains comparatively rigid, while the peduncle and caudal fin execute large‑amplitude oscillations \cite{khalid2021larger, sfakiotakis2002review}. These differences in \edt{profile of the undulation} amplitude, effective wavelength, and \edt{kinematics of the} caudal fin strongly influence power consumption and attainable swimming speeds. \edt{Particularly}, intermittent burst‑and‑coast locomotion is widely observed in both gait families and plays a central role in energy management \cite{liu2025intermittent}. During the burst phase, active undulations produce large circulation and high instantaneous thrust. \edt{During} the coast phase, the body becomes effectively rigid and straight\edt{,} and it moves within the previously generated momentum field. Some \edt{previous} studies \edt{suggested} that alternating between these phases \edt{reduced} average muscular effort and overall metabolic cost relative to strictly continuous undulation \cite{wu2011fish, dai2018intermittent, liu2025intermittent}, with \edt{some energy saved due to the} inertial wake generated in \edt{the} burst phase \cite{li2023intermittent, yang2024hydrodynamics}. Continuous gaits tend to produce regular shedding \edt{coherent flow structures}, whereas burst‑and‑coast strategies introduce transient amplification of vortex structures as the body \edt{is} alternately driven \edt{in the two phases}. These differences are expressed directly in \edt{temporal profiles of} thrust, drag, and side‑force. 


\edt{Most of the previous investigations focused on examining the vortex dynamics and its influence on the hydrodynamic performance of fish, performing continuous undulations \cite{borazjani2008numerical, liu2017computational, khalid2021anguilliform, khalid2021larger, fardi2025characterizing}. Besides, there exist a few studies that explored the fluid dynamics and quantifying the associated benefits of intermittent swimming of two-dimensional undulating bodies in terms of energy expenditures \cite{dai2018intermittent, akoz2018unsteady, li2023intermittent, yang2024hydrodynamics, liu2025intermittent}. These studies suffer from limitations, such as the utilization of inviscid flow based modeling techniques, and simple two-dimensional foil-like geometric bodies, which do not capture the complex vortex-body and vortex-vortex interactions, happening around real marine organisms. According to the authors' knowledge, there exists no study in the current literature that explains the three-dimensional vortex dynamics and its role on production of instantaneous hydrodynamic forces of biological swimmers. Ashraf et al. \cite{ashraf2020burst} performed experiments and three-dimensional simulations for flows around a rummy nose tetra fish (Hemigrammus bleheri) to only examine the power requirements for various swimming modes, including burst-and-coast motion. With more interest arising in the marine engineering community to develop bio-inspired underwater robots, it is an important knowledge gap that our present study addresses here for two important classes of biological swimmers, including carangiform and anguilliform. The hierarchical relation between the duty cycle (defined as the fraction of the cycle spent in active undulation) and energy consumption by undulating swimmers was  previously explained by Dai et al. \cite{dai2018intermittent} for two-dimensional elastic bodies and Liu et al. \cite{liu2025intermittent} for an experimental fish-inspired robot. Nevertheless, the combined effects of duty cycle and undulatory kinematic modes of swimmers along with their morphologies for different Strouhal numbers on hydrodynamics and forces are not elucidated earlier. 

To investigate these important hydrodynamic aspects for biological swimmers and their bio-inspired counterpart robotic systems, we perform high-fidelity three-dimensional simulations around two different marine swimmers, including a Jack fish (\textit{Crevalle Jack} or \textit{Caranx Hippos}) and an American eel (\textit{Anguilla Rostrata}), which are representatives of carangiform and anguilliform classes \cite{khalid2021anguilliform, khalid2021larger}, respectively. The primary aim of our present work is to examine the unsteady flow dynamics around the two physiologically and kinematically very different swimmers, while they perform the burst-and-coast motion as well as continuous undulations under varying Strouhal number. Our computational simulations using the in-house fully parallelized sharp-interface immersed-boundary solver, VorteXdyn, reveal intriguing similarities in the hydrodynamics of the two swimmers. Our further analyses also unveils key systematic variations in the wake dynamics of these swimmers with a consistent changes in duty cycle through their respective undulatory kinematic profiles. }

\section{Computational Methodology}
\label{sec:Num_Method}
In this study, we simulate three-dimensional flow around real geometries of two fish species, including the American eel and the Jack Fish\edt{, for which the morphogical} data for these swimmers is \edt{adopted from the work previously done by Khalid et al.} \cite{khalid2021anguilliform, khalid2021larger}. The Jack Fish, consisting of trunk and caudal fin sections, has \edt{a} total height and width of $0.3021L$ and $0.1481L$, respectively, \edt{with} a body area of 0.4479$L^2$. \edt{Here, $L$ presents the body-length of a swimmer.} Its surface mesh comprises $22,712$ triangular elements for the trunk and $2,560$ for the caudal fin. The eel \edt{has a slender geometry}, with \edt{a} total height and width of 0.0884$L$ and 0.0605$L$, \edt{respectively, with the} surface area of 0.1739$L^2$, \edt{which is} discretized using $33,112$ triangular elements. 

\begin{figure*}[htbp]
\centering
\includegraphics[width=6in]{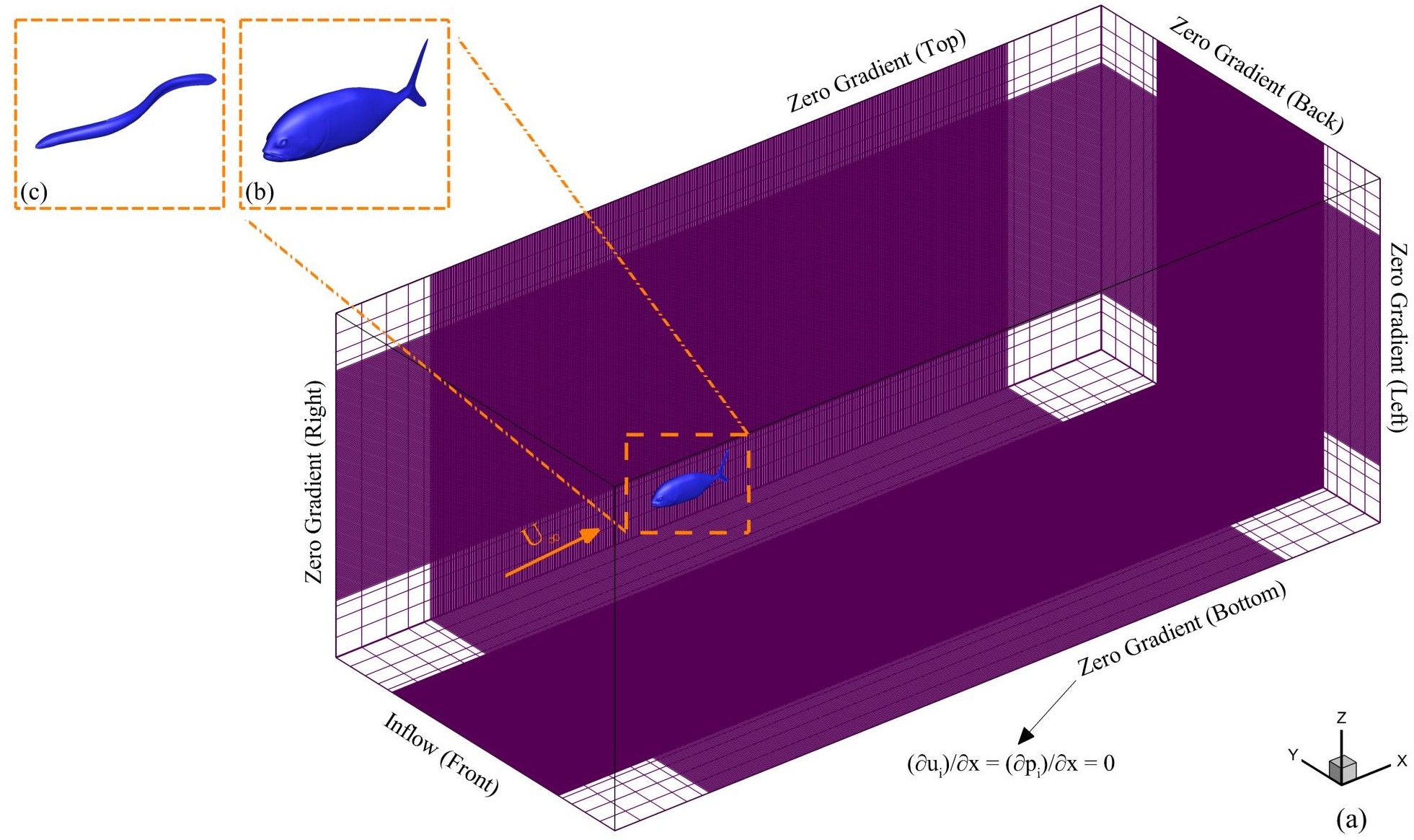}
\caption{\small \edt{(a) Flow domain with specifc boundary conditions on its sides and the mesh regions around a swimmer's body, (b) physiology of a Jack Fish, and (c) geometrical features of an eel}}
\label{fig:domain}
\end{figure*}


\begin{figure*}[htbp]
\centering
\includegraphics[width=5in]{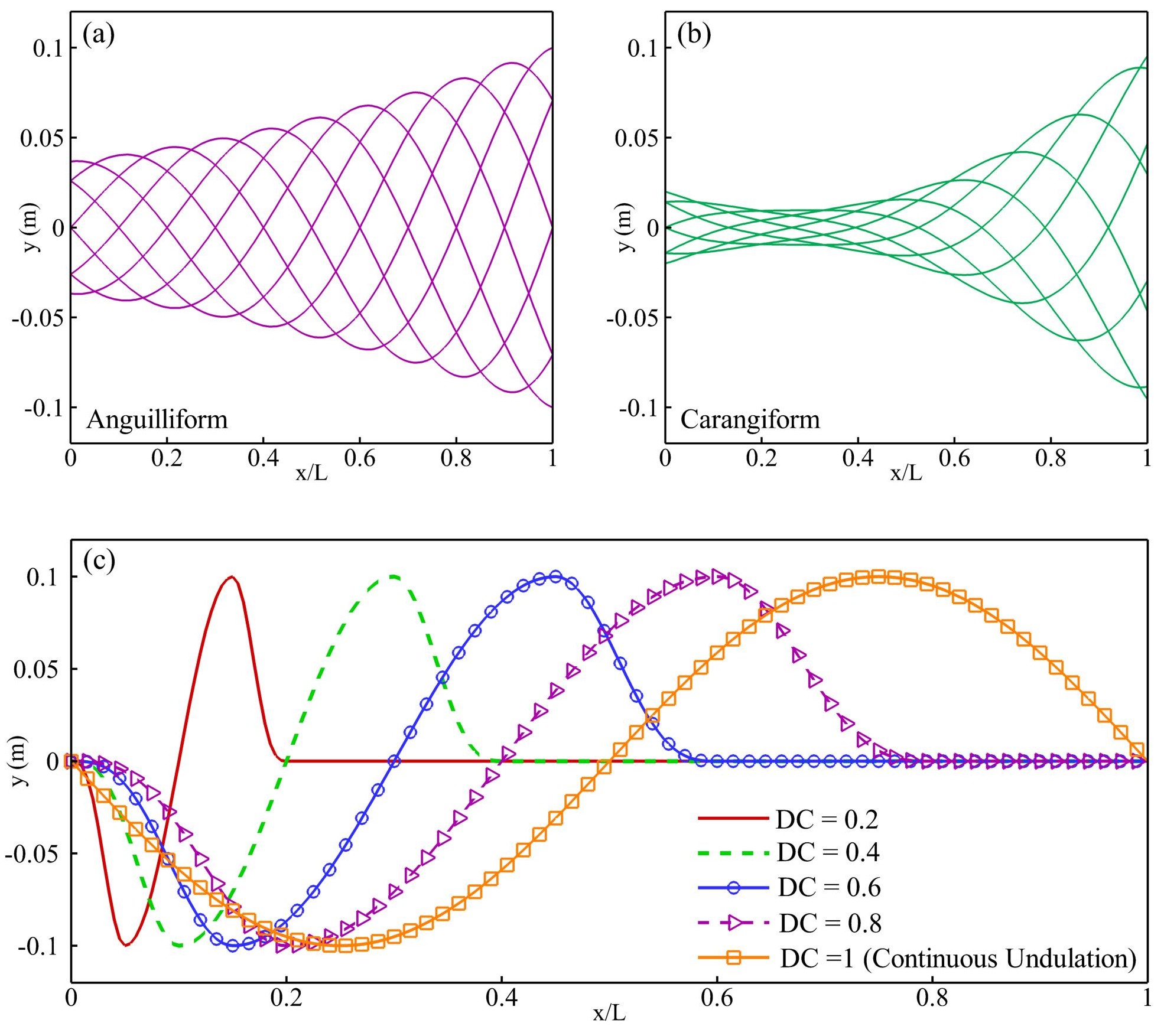}
\caption{\small \edt{Undulatory kinematic profiles for the (a) anguilliform and (b) carangiform modes, and (c) the kinematic profiles of the swimmer's tail during burst-and-coast motion with different duty cycles}}
\label{fig:profile}
\end{figure*}

Each swimmer's body kinematics follow a wavy motion defined by a traveling wave function combined with a temporal smoothing function to transition between \edt{the} burst and coast phases. This formulation is based on the approach proposed by \edt{Yang et al.} \cite{yang2024hydrodynamics}. \edt{Equation~\ref{eq:burst_coast} describes the the instantaneous local undulatory kinematics on each point on a swimmer's body.} 

\begin{equation}
y(x,t) =
\begin{cases}
S(t) A_m(x) \sin\!\left[ 2\pi\!\left( \dfrac{x}{\lambda} - \dfrac{t}{T_s} \right) - \dfrac{2\pi}{\lambda} \right], & 0 \le t \le T_s, \\
0, & T_s < t \le T,
\end{cases}
\label{eq:burst_coast}
\end{equation}

\noindent \edt{where the oscillation amplitude $A_m$ as a function of the location of a point on the body is provided below using second-order polynomials for the carangiform motion \cite{videler1984fast, khalid2021larger}:

\begin{align}
    A\left(\frac{x}{L}\right) &= 0.02 - 0.0825\left(\frac{x}{L}\right) + 0.1625\left(\frac{x}{L}\right)^2; \quad 0 < \frac{x}{L} < 1
\label{eq:carangiform}
\end{align}

For the anguilliform motion, $A_m$ is defined as \cite{khalid2021anguilliform}:

\begin{align}
    A\left(\frac{x}{L}\right) &= 0.0367 + 0.0323\left(\frac{x}{L}\right) + 0.0310\left(\frac{x}{L}\right)^2; \quad 0 < \frac{x}{L} < 1
\label{eq:amp}
\end{align}

A cosine-based smoothing function in Eq.~\ref{eq:smooth}} modulates the undulatory kinematics during \edt{the burst phase}: 

\begin{equation}
S(t) =
\begin{cases}
0.5 \left[ 1 - \cos\!\left( \dfrac{4\pi t}{T_s} \right) \right], & 0 \le t \le 0.25T_s, \\[8pt]
1, & 0.25T_s < t \le 0.75T_s, \\[8pt]
0.5 \left[ 1 - \cos\!\left( \dfrac{4\pi t}{T_s} \right) \right], & 0.75T_s < t \le T_s.
\end{cases}
\label{eq:smooth}
\end{equation}

In the equations above, $t$ is the time, $\lambda$ \edt{denotes} the wavelength of the traveling wave \edt{along a swimmer's body}, $T_s$ \edt{shows} the time for the \edt{burst phase in one full burst-and-coast cycle, for which $T$ is the total period. Therefore,} the non-dimensional quantity, duty cycle ($\mbox{DC}$), \edt{represents} the fraction of \edt{the} time spent \edt{for active undulation} within a swimming cycle \cite{akoz2018unsteady}. \edt{It is defined below:

\begin{equation}
DC = \frac{\text{Time for active undulation during burst}}{\text{total time for one burst-and-coast cycle}} = \frac{T_s}{T}
\label{eq:DC}
\end{equation}
}

The governing mathematical model for fluid flows is based on the following non-dimensional forms of the continuity and incompressible Navier-Stokes equations \cite{farooq2025accurate,mittal2008versatile}: 

\begin{equation}
\begin{aligned}
\frac{\partial u_j}{\partial x_j} &= 0, \\
\frac{\partial u_i}{\partial t} + u_j \frac{\partial u_i}{\partial x_j}
&= -\frac{1}{\rho} \frac{\partial p}{\partial x_i}
+ \frac{1}{Re} \frac{\partial^2 u_i}{\partial x_j \partial x_j}
+ f_b
\end{aligned}
\end{equation}

\noindent where ${i, j} = {1,2,3}$, $x_i$ represents the Cartesian directions, $u_i$ denotes the Cartesian components of \edt{velocity of the fluid}, $p$ is the pressure, and $\mbox{Re}$ is the Reynolds number. We define it as $\mbox{Re}={U_\infty}{L}/\nu$, where $U_\infty$ shows the free-stream velocity, and $\nu$ is kinematic viscosity of the fluid. In this formulation, $f_b$ is a discrete forcing term that enables a sharp representation of the immersed boundary \cite{mittal2008versatile, farooq2025accurate}.

We solve these equations using our in-house sharp-interface immersed boundary method (IBM) based solver, \textbf{VorteXdyn}, to accurately capture the interaction between the fluid and the complex \edt{geometry of the swimmers, containing volumetric parts and membranous elements} \cite{farooq2025accurate, fardi2025characterizing}. Spatial derivatives are discretized using \edt{the} second-order central difference \edt{and Quadratic Upstream Interpolation for Convective Kinematics (QUICK) schemes for the diffusion and convective terms, respectively. Time marching in this solver is carried out} using \edt{the} fractional-step method that ensures second-order accuracy. Boundary conditions are imposed such that a Dirichlet condition is used for the inflow, while Neumann conditions are enforced at all other boundaries of the domain, as also explained in Fig.~\ref{fig:domain}. No-slip boundary conditions are applied on the swimmer’s body using \edt{the ghost cell-based approach \cite{farooq2025accurate}. Further details on the verification and validation of the solver can be found in the studies by Farooq et al. \cite{farooq2025accurate} and Kamran et al. \cite{kamran2025role}} to ensure the accuracy and reliability of our solver.

\section{Results and Discussion}
\label{sec:results}
\edt{The primary objective of this work is to examine the three-dimensional vortex dynamics around swimmers and in their wakes and to focus on how hydrodynamic forces are influenced by these flow mechanisms while they perform burst-and-coast motion with different duty cycles.} The governing flow and kinematic parameters used here are summarized in \edt{Table~\ref{tab:parameters}. The simulations here are conducted at $\mbox{Re}=3,000$, lying within a transitional flow regime \cite{borazjani2008numerical}. Previously, Liu et al. \cite{liu2017computational} and Zhong et al. \cite{zhong2019dorsal} concluded that primary flow features around oscillating bodies remained the same with varying Reynolds number. Both carangiform and anguilliform} waveforms are prescribed \edt{over the respective swimmers} for Strouhal numbers ($\mbox{St}$) of $0.30$ and $0.40$, consistent with the typical swimming patterns of marine animals \cite{fish2020advantages, liu2017computational}. \edt{Here, Strouhal number is defined as $\mbox{St}={2}\edt{{A_\circ}}{f}/U_\infty$, where $f$ presents the undulation frequency, and $A_\circ$ is the maximum one-sided oscillation amplitude of the swimmer's tail.} It is important to note that the undulation wavelength is set to $\lambda/L = 0.80$ for the anguilliform kinematic mode \cite{khalid2021anguilliform} and $\lambda/L = 1.05$ for the carangiform mode \cite{khalid2021larger, liu2017computational}.

\begin{table}[h!]
\centering
\caption{ \small Specifications of the governing parameters}
\begin{tabular}{ll}
\hline
\textbf{Parameters} & \textbf{Specifications} \\ 
\hline
Swimmers & Eel and Jack Fish \\[4pt]
Undulatory kinematics & Anguilliform and Carangiform \\[4pt]
Motion & Burst and Coast ($\mbox{DC}=0.2-0.8$) \& Continuous ($\mbox{DC}=1$)\\[4pt]
$\mbox{Re}$ & $3000$ \\[4pt]
$\mbox{St}$ & $0.30$ and $0.40$ \\[4pt]
$\lambda/L$ & $0.80$ (anguilliform) and $1.05$ (carangiform) \\ 
\hline
\end{tabular}
\label{tab:parameters}
\end{table}

\begin{figure}[htbp]
\centering
\includegraphics[width=6in]{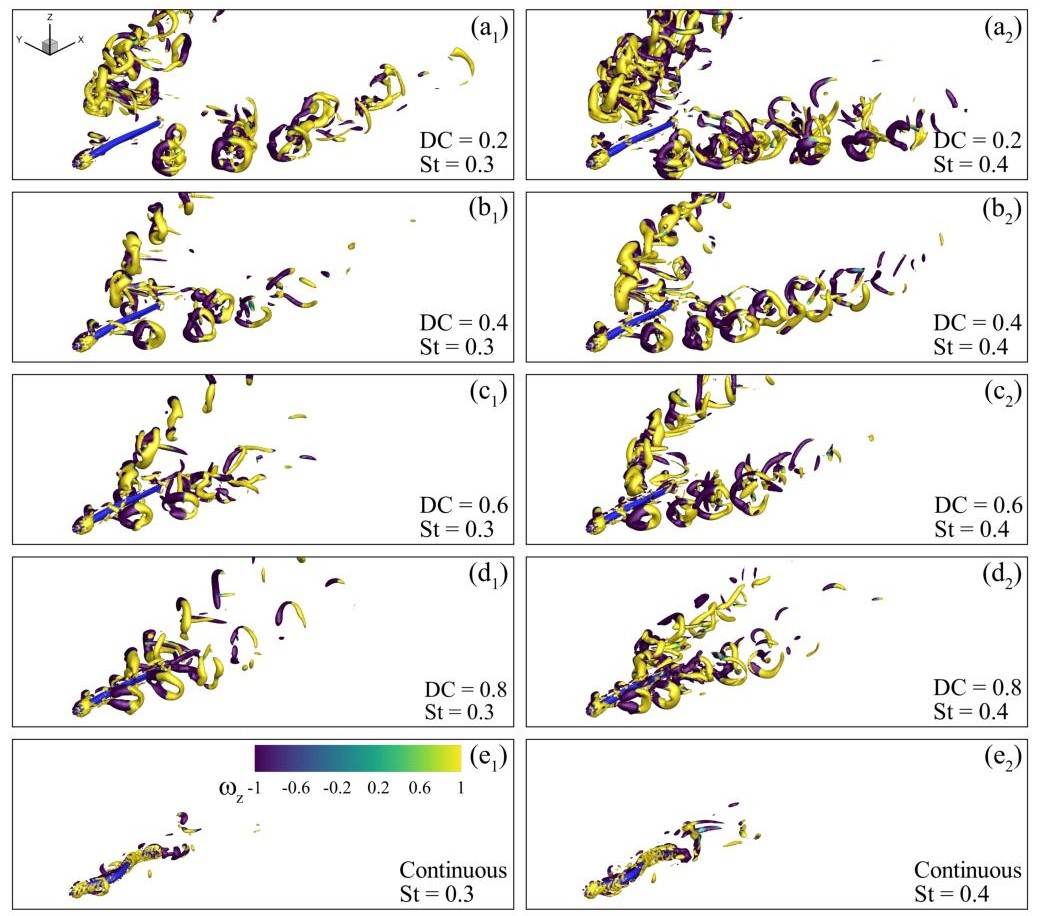}
\caption{\small Vortex structures around the anguilliform swimmer at the end of the undulation cycle in 5 modes for ($a_1$ to $e_1$) $\mbox{St} = 0.3$ and ($a_2$ to $e_2$) $\mbox{St} = 0.4$.}
\label{figure1}
\end{figure}

We start the discussion by \edt{explaining the} development of \edt{three-dimensional coherent flow features} around \edt{the two swimmers for different undulatory kinematic profiles}. Figure~\ref{figure1} provides visualizations of \edt{vortices} around the anguilliform swimmer at the \edt{end of the undulation cycle in each case} of $\mbox{DC}$ and Strouhal number \edt{considered in this work}. We notice that the swimmer forms bow-like wakes for \edt{$\mbox{DC}=0.2-0.8$} (see \edt{Figs.}~\ref{figure1}$a_1$ to \edt{\ref{figure1}$d_1$ for $\mbox{St}=0.30$} and \edt{Figs.}~\ref{figure1}$a_2$ to \edt{\ref{figure1}$d_2$ for $\mbox{St}=0.40$}). These wakes are constituted by two well-separated rows of vortices on either side of the body, angled outward with the row on the right that consistently \edt{gets formed} at a slightly higher angle from the body. This asymmetry arises from the phase of the body’s oscillation and can get inverted laterally by introducing a phase of $\pi$ in the undulatory profile. Close to the body, shear layers roll up into hairpin\edt{-like} vortices, which \edt{get detached from the body} and combine with trailing-edge vortices to form \edt{ring-like} structures in the wake. On the \edt{anterior side} of the swimmer, there is always a \edt{shear} layer due to the interaction between the flow and the body. As the duty cycle increases, these dual vortex rows shift closer to the swimmer’s centerline \edt{(or that of the wake)} while fewer vortices are shed per cycle, though the asymmetry in angles remains. Also, the vortex shedding \edt{process} lasts less by increasing $\mbox{DC}$. 

\begin{figure}[htbp]
\centering
\includegraphics[width=6in]{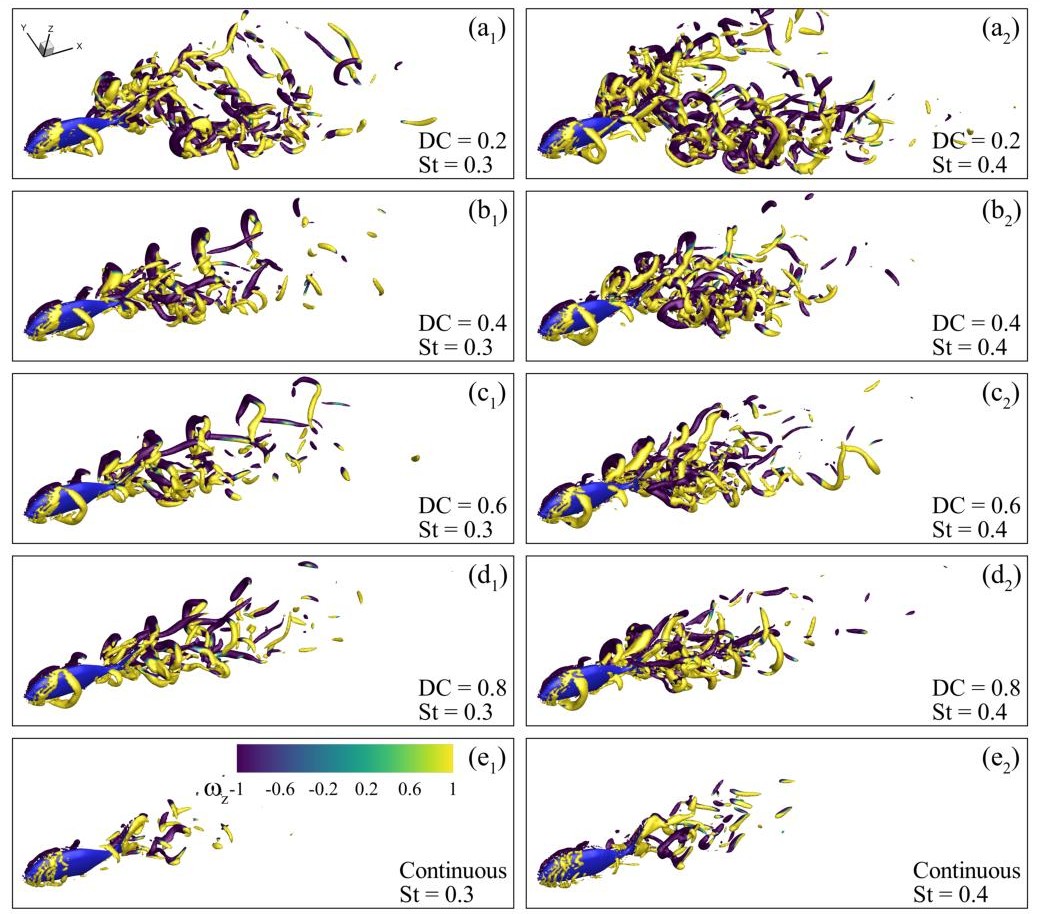}
\caption{\small Vortex structures around the carangiform swimmer at the end of the undulation cycle in 5 modes for ($a_1$ to $e_1$) $\mbox{St} = 0.3$ and ($a_2$ to $e_2$) $\mbox{St} = 0.4$.}
\label{figure2}
\end{figure}

Turning to the carangiform swimmer, \edt{visualization in} Fig.~\ref{figure2} \edt{show} the evolution of the wake at the final time instant \edt{for all the values of} $\mbox{DC}$ and both Strouhal numbers. For $\mbox{DC}=0.2-0.8$, the wake organizes into a bow-like pair of vortex streets \edt{that are similar to those of} the anguilliform swimmer (\edt{Figs}.~\ref{figure2}$a_1$ to \ref{figure2}$d_1$ for $\mbox{St}=0.30$ and Figs.~\ref{figure2}$a_2$ to \ref{figure2}$d_2$ for $\mbox{St}=0.40$). Two rows detached from the body and lean outward, with the \edt{left_sided} row persistently steeper in angle. Immediately adjacent to the surface, shear layers roll up into hairpins, \edt{which then} couple with trailing-edge vortices to \edt{be} convected downstream. A thin leading-edge \edt{shear} layer is visible along the \edt{anterior} part of the body. With \edt{an} increasing $\mbox{DC}$, the two rows draw toward the centerline and the number of structures \edt{produced} per cycle drops, while the angular asymmetry remains evident. \edt{Concurrently, the vortices get dissipated in the wake more quickly, as clearly presented in Figs.}~\ref{figure2}$a_1$ to \ref{figure2}$d_1$ as an example). Relative to \edt{wakes} of the anguilliform \edt{swimmer} in Fig.~\ref{figure1}, the \edt{flow features around the} carangiform \edt{swimmer} in Fig.~\ref{figure2} show a busier near-body roll-up, producing denser rows and larger, longer-lived \edt{vortex structures}.

\begin{figure}[htbp]
\centering
\includegraphics[width=5in]{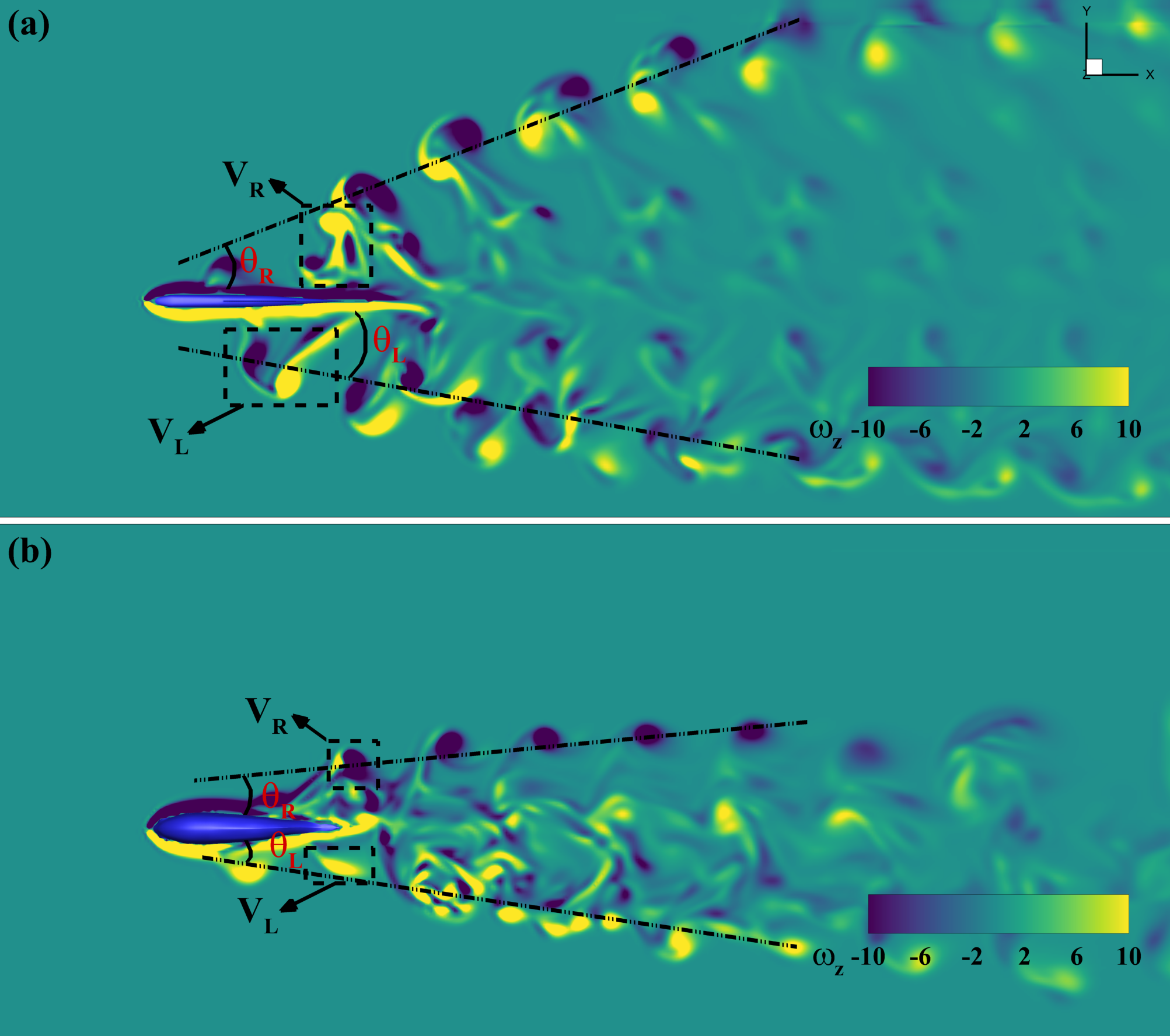}
\caption{\small Vorticity $(\omega_z)$ at $z = 0$ \edt{plane} for $\mbox{DC} = 0.4$ at $\mbox{St} = 0.3$ (a) anguilliform swimmer, (b) carangiform swimmer.}
\label{figure3}
\end{figure}

\begin{figure}[htbp]
\centering
\includegraphics[width=6.5in]{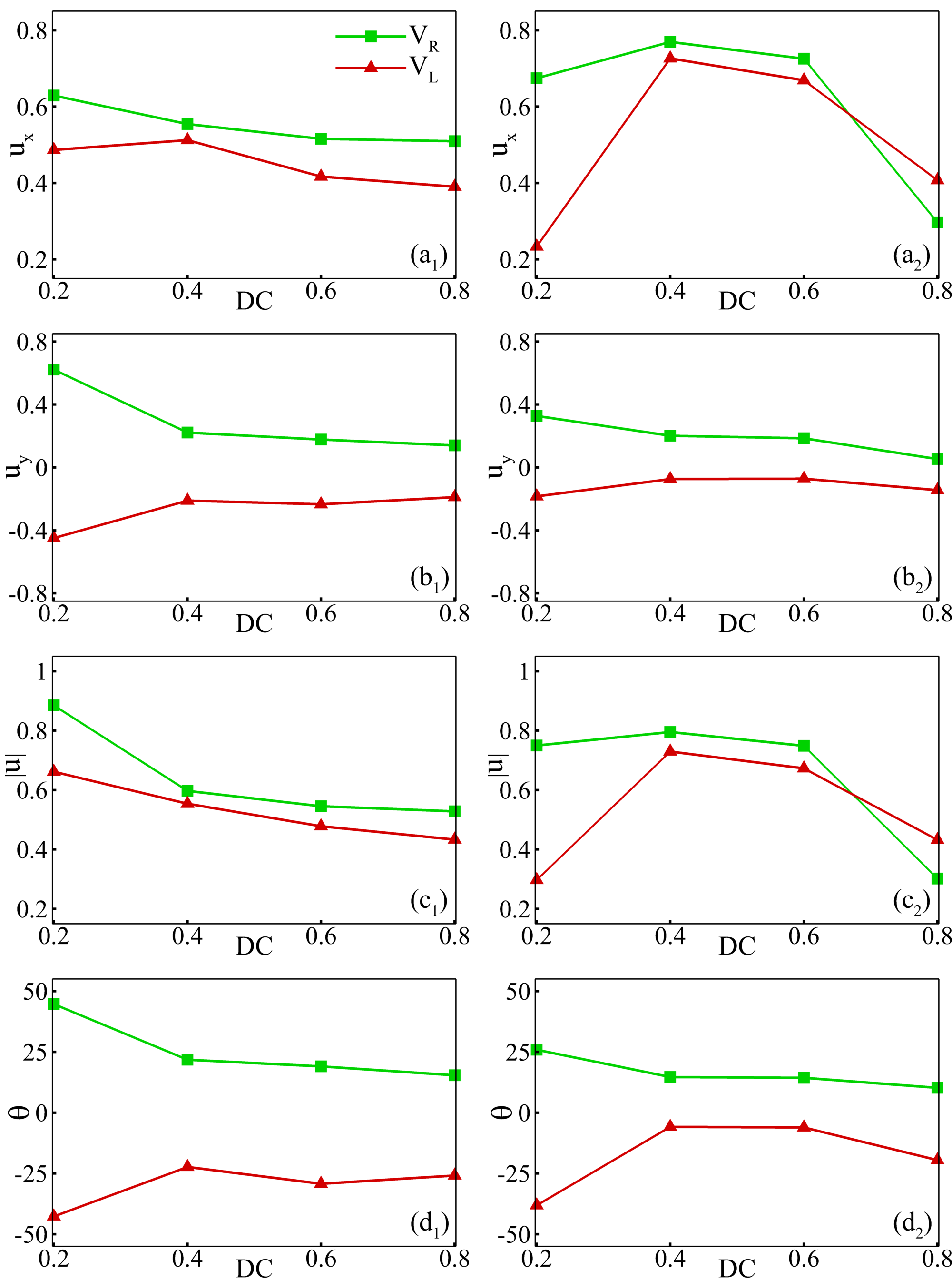}
\caption{\small \edt{Cycle-averaged kinematic parameters of vortices} on \edt{the} right and left side of \edt{a swimmer's body} versus $\mbox{DC}$ at $\mbox{St} = 0.3$ for eel \edt{(left column)} and Jack Fish \edt{(right column), where} (a) $u_x$, (b) $u_y$, (c) $|u|$, (d) $\theta_R$, \edt{and} $\theta_L$.}
\label{figure4}
\end{figure}

\edt{The first important question arise here about the decreasing angle of the bow-shaped wakes for the two swimmers with an increasing values of $\mbox{DC}$. To understand this behavior}, we show the \edt{out-of-plane component of vorticity $\omega_z$ in the} mid plane ($z=0$) \edt{of the flow field in Fig.~\ref{figure3} and} label the right\edt{-} and left\edt{-sided} vortices \edt{along with} the bow angles $\theta_R$, $\theta_L$ at $\mbox{DC} = 0.4$ and $\mbox{St} = 0.3$. These {plots explains the relevant notations} and links the two vortex rows to the \edt{swimmers' straight bodies}. We then track the vortices for the eel and present their streamwise speed ($u_x$), cross-stream speed ($u_y$), total speed $|u|$, and the \edt{bow} angles $\theta_R$ and $\theta_L$ \edt{in Fig.~\ref{figure4}} as $\mbox{DC}$ \edt{increases} from $0.2$ to $0.8$. \edt{These velocities for the vortices $V_R$ and $V_L$ are determined through tracking their centers using the algorithm presented by Khalid et al. \cite{khalid2020flow} and computing the time-averaged values. The plots in Figs.~\ref{figure4}a1 and \ref{figure4}a2 demonstrate that} both vortices \edt{mostly} slow \edt{down} in the streamwise direction\edt{, as $\mbox{DC}$ rises, whereas} their cross-stream \edt{velocity components} moves toward zero on each side (see Figs.~\ref{figure4}b1 and \ref{figure4}b2). The total speed \edt{also} drops for $\mbox{DC} > 0.40$ in Figs.~\ref{figure4}c1 and \ref{figure4}c2. Most importantly, Figs.~\ref{figure4}d1 and \ref{figure4}d2 clearly exhibit that $\theta_R$ and $\theta_L$ fall toward $0^\circ$, and the vortex rows \edt{gets deflected} inward. This collapse in $\theta$ explains \edt{how} the bow angle decreases with $\mbox{DC}$. \edt{The salient difference in these metrics for the Jack Fish in comparison to those for the eel include: (i) the overall trends for $u_x$, $u_y$, $|u|$, $\theta_R$, and $\theta_L$ are similar to those of the eel but their change with respect to $\mbox{DC}$ is relatively smaller; (ii) the cross-stream speeds of the two vortices approach each other with less steepness with the increasing $\mbox{DC}$ in the case of the Jack Fish, (iii) the carangiform swimmer's motion is of a large amplitude at the tail and the anterior part of the body performs significantly small-amplitude undulation. Due to this reason, the caudal fin sheds the vortices more axially right from the beginning of the undulation cycle, so a narrower wake is formed. It may leave less room for further narrowing of the wake as we increase $\mbox{DC}$. Moreover, caudal fin of the Jack Fish holds the leading-edge vortex on its surface for a longer time and guides pairing between the vortices near the trailing edge \cite{borazjani2008numerical}. It helps stabilize the near-wake and reduces its sensitivity to duty cycle. These important features make the Jack Fish respond less to $\mbox{DC}$ than the eel, which has a slender axisymmetric-like body performing undulations with large amplitudes all along its body.}




In \edt{Figs}.~\ref{figure4}$a_1$ to $d_1$, we also see a clear step in $u_y$ when we move from \edt{a} low to higher $\mbox{DC}$. \edt{For a} low $\mbox{DC}$, the short burst kicks new vortex off the body before \edt{other previously shed vortices can} pair, so the vortices leave with larger cross-stream speed and \edt{form} a wide bow. When we extend the \edt{time for the} burst \edt{phase} at \edt{a} higher $\mbox{DC}$, \edt{the vortices do not get detached from the body at an early stage and traverse along it to do so at a more posterior position. This delay in the detachment of vortices from the swimmer's body help the vortices on the left and right sides of the body become more symmetric. A metric to examine the influence of the formation of the asymmetric vortices on the two sides of the swimmer could be the side force ($C_y$). Figure~\ref{figure6} provides the plots of the time-averaged $C_y$ for different values of $\mbox{DC}$ and $\mbox{St}$ for both swimmers.}


\begin{figure*}[htbp]
\centering
\includegraphics[width=6.5in]{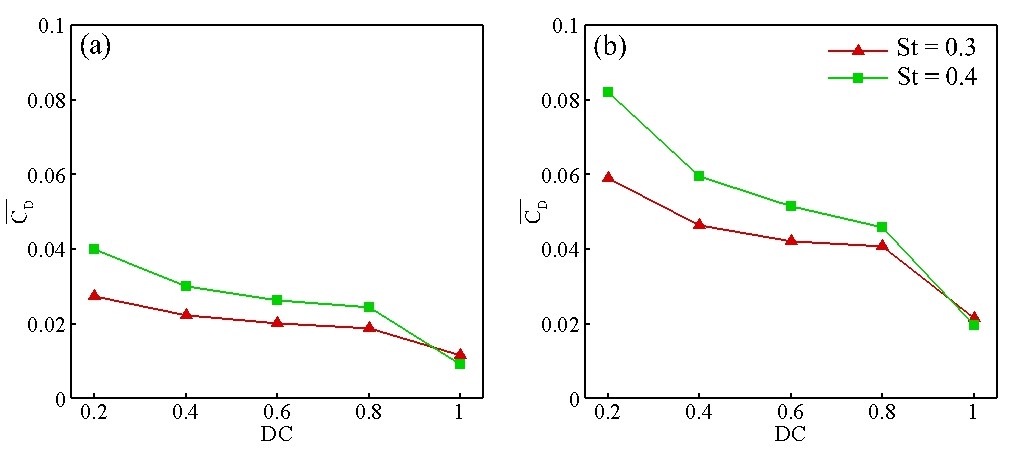}
\caption{\small Time-averaged drag coefficient for (a) anguilliform swimmer and (b) carangiform swimmer.}
\label{figure5}
\end{figure*}

\begin{figure}[htbp]
\centering
\includegraphics[width=6.5in]{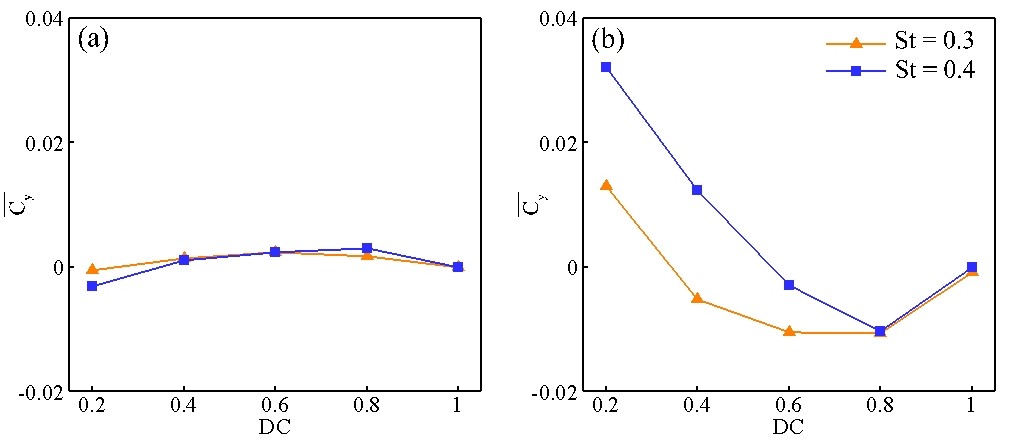}
\caption{\small Time-averaged side force coefficient for \edt{the} (a) anguilliform swimmer and (b) carangiform swimmer.}
\label{figure6}
\end{figure}

\edt{Now, we explain the hydrodynamic forces experienced by the swimmers under different kinematic conditions.} Figure~\ref{figure5} shows the mean drag \edt{coefficient ($\bar{C_D}$), averaged} over the last five cycles for the eel (anguilliform) and the Jack Fish (carangiform). In Fig.~\ref{figure5}$a$ (eel), drag steadily drops as the duty cycle goes from $0.2$ to $1.0$ for both $\mbox{St}=0.3$ and $0.4$. \edt{These results from the present study corroborate with the findings of Yang et al. \cite{yang2024hydrodynamics}, where they reported increase in an undulating swimmer's speed through their two-dimensional flow simulations.} At \edt{a} low $\mbox{DC}$, the higher Strouhal number makes the bursts stronger and more abrupt, so the flow around the body is more disturbed and the average drag is higher. As $\mbox{DC}$ increases, the coast phase gives the flow time to settle, the wake points straighter downstream, and drag keeps reducing for both swimmers (see Fig.~\ref{figure5}). \edt{Quite interestingly, $\bar{C_D}$ is greater for $\mbox{St}=0.40$ than that at $\mbox{St}=0.30$ for the intermittent swimming (for all $\mbox{DC}<1.0$). Generally, a continuously undulating swimmer's hydrodynamic performance improves in terms of reducing drag, or equivalently increasing thrust, with increasing $\mbox{St}$ \cite{liu2017computational, khalid2021anguilliform}, as also evident for $\mbox{DC}=1$ for the two swimmers in Figs.~\ref{figure5}. However, this trend for the influence of $\mbox{St}$ on both swimmers' performance is reversed for the intermittent swimming modes ($\mbox{DC}<1$). }


Figure~\ref{figure6} shows the time-averaged side force coefficient ($\bar{C_y}$) for both eel and Jack Fish. For the eel (Fig.~\ref{figure6}$a$), it stays close to zero for all $\mbox{DC}$ and both Strouhal numbers, with a gentle bump around $\mbox{DC} = 0.8$. This small bump suggests a bit of leftover asymmetry from the mid-body motion at a high $\mbox{DC}$, but it is mild and goes away again at $\mbox{DC} = 1$. The side force coefficient of Jack Fish (Fig.~\ref{figure6}$b$) varies more. At \edt{low values of} $\mbox{DC}$, the side force is large, especially at $\mbox{St} = 0.4$. \edt{The potential reason could be} the stronger bursts \edt{pushing} the wake sideways, and the jet does not point straight back. When we increase $\mbox{DC}$, the timing of the vortices \edt{traversing to the posterior part of the swimmer} and \edt{motion of the caudal fin help keep the wake straight} and the side force drops fast. It reaches a clear minimum around $\mbox{DC} = 0.7–0.8$ and then climbs back toward zero when we reach continuous motion. 


\begin{figure*}[htbp]
\centering
\includegraphics[width=6.5in]{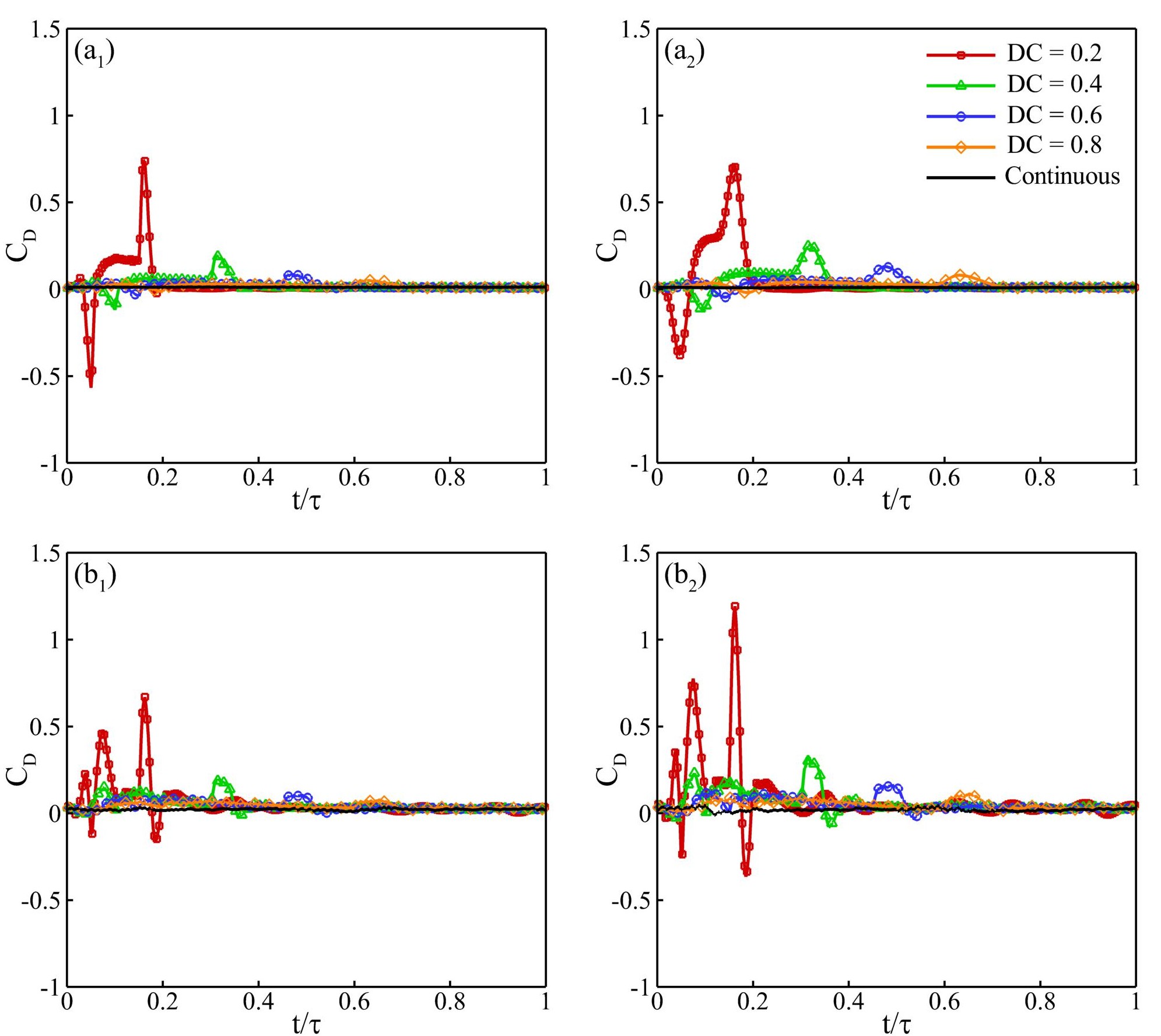}
\caption{\small Instantaneous drag force over one cycle in the anguilliform swimmer at ($a_1$) $\mbox{St} = 0.3$ and ($a_2$) $\mbox{St} = 0.4$, and in the canrangiform swimmer at ($b_1$) $\mbox{St} = 0.3$ and ($b_2$) $\mbox{St} = 0.4$.}
\label{figure7}
\end{figure*}

Figure~\ref{figure7} shows the instantaneous drag coefficient ($C_D$) for \edt{different values of} duty cycles at both Strouhal numbers for the eel and Jack Fish. We \edt{consider} positive $C_D$ as drag and negative $C_D$ as thrust. \edt{For the} anguilliform \edt{swimmer} (Fig.~\ref{figure7}$a_1$ and \ref{figure7}$a_2$) at $\mbox{DC} = 1$ (continuous), $C_D$ stays close to zero over the whole cycle, and only small oscillations are present. \edt{Importantly, this condition indicates the free propulsion of the swimmer, where thrust and drag are balanced, and the swimmer performs steady swimming.} In burst and coast \edt{motion}, the response becomes strongly transient. Right after the burst starts, we see a short-thrust dip (negative $C_D$) as the body accelerates the fluid, followed by a large drag peak while the body slows down. As $\mbox{DC}$ increases from $0.2$ to $0.8$, both the thrust dip and the drag peak shrink, and the curve moves back toward the flat continuous baseline. This trend \edt{persists} at both $\mbox{St} = 0.30$ and $0.40$.  Overall, $\mbox{DC}$ \edt{appears to be a strong controlling factor to determine} how impulsive the \edt{temporal profile of the streamwise force} looks. \edt{On the other hand,} the Jack Fish \edt{exhibits} slightly larger oscillations than the eel but with no large excursions for $\mbox{DC} = 1$  \edt{(see Figs.~\ref{figure7}$b_1$ and \ref{figure7}$b_2$). For the intermittent kinematics}, the burst phase is dominated by drag with short thrust\edt{-producing periods}. At \edt{a} low $\mbox{DC}$, the waveform \edt{of the streamwise force} often shows three drag peaks and two thrust dips \edt{when the swimmer performs bursting motion}. \edt{These observations might be caused} from the initial acceleration and a brief rebound as the tail finishes the burst. As $\mbox{DC}$ increases, all peaks drop in magnitude \edt{get further delayed in the kinematic cycle}. The effect is strongest at $\mbox{St} = 0.4$, where the initial burst is \edt{more pronounced} and produces the largest peaks at low \edt{values of} $\mbox{DC}$.



\begin{figure}[htbp]
\centering
\includegraphics[width=6.5in]{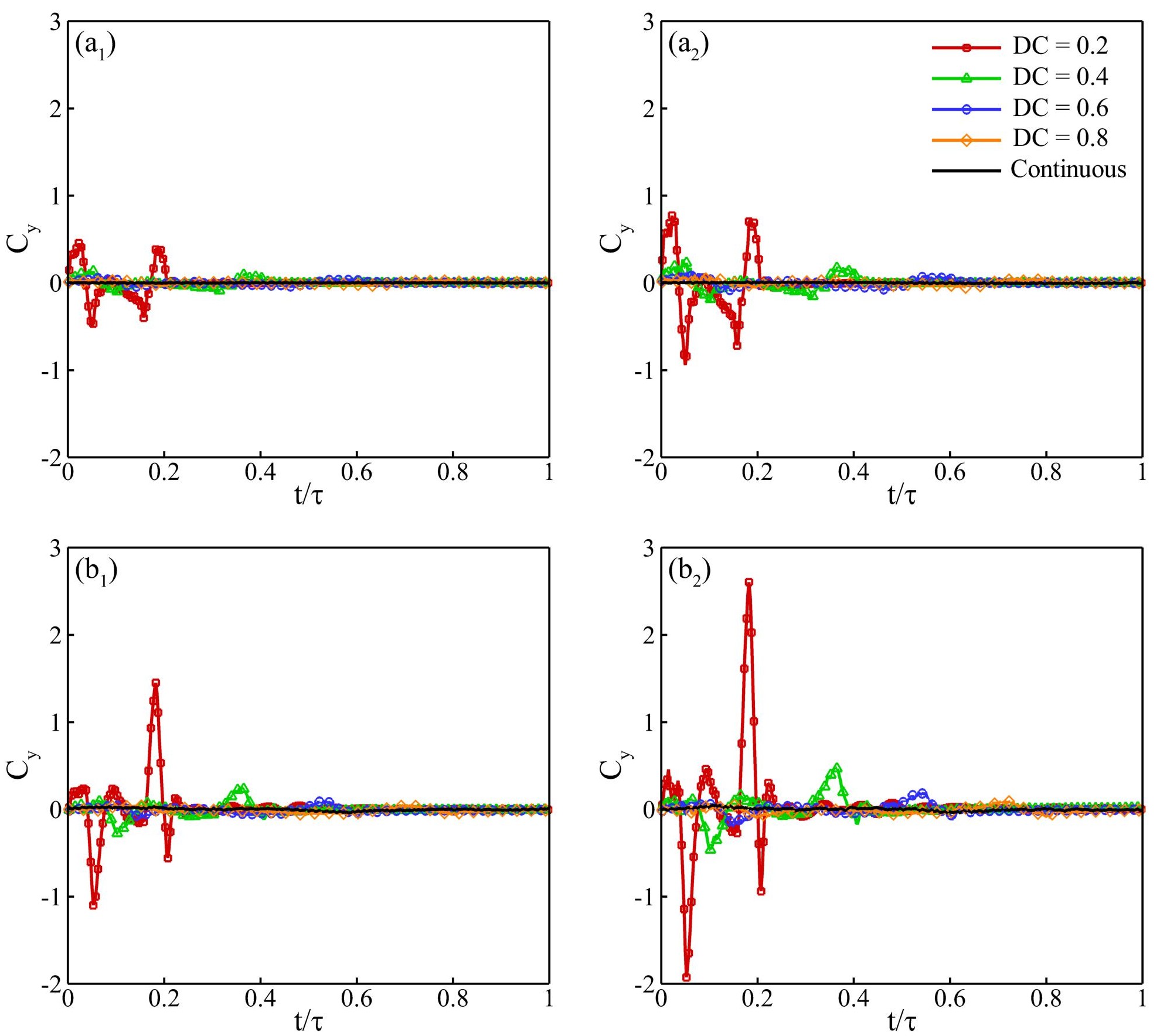}
\caption{\small Instantaneous side force over one cycle \edt{for} the anguilliform swimmer at ($a_1$) $\mbox{St} = 0.3$ and ($a_2$) $\mbox{St} = 0.4$, and \edt{for} the carangiform swimmer at ($b_1$) $\mbox{St} = 0.3$ and ($b_2$) $\mbox{St} = 0.4$.}
\label{figure8}
\end{figure}

Figure~\ref{figure8} plots the instantaneous side-force coefficient ($C_y$) for \edt{different duty cycles and Strouhal numbers for} both swimmers. Positive and negative values indicate force to the right and left, respectively. For the anguilliform \edt{swimmer} (eel) in Fig.~\ref{figure8}$a_1$ and \ref{figure8}$a_2$, \edt{${C_y}$ for}  the continuous case ($\mbox{DC} = 1$) is almost flat, showing only slight oscillations. \edt{For burst and coast motion of both swimmers}, short-lived pulses \edt{in $C_y$} appear during the burst window and then fade to near-zero in the coast \edt{phase}. The pulses are fairly symmetric, a push in one direction is followed by a similar push in the other. As $\mbox{St}$ increases from $0.30$ to $0.40$, those pulses in the burst phase become \edt{more significant. Comparatively, the instant increments and decrements in $C_y$ are larger for the Jack Fish in Figs.~\ref{figure8}$b_1$ and \ref{figure8}$b_2$}. At \edt{a} low $\mbox{DC}$, the burst motion often shows a cluster of sharp peaks\edt{, which are likely to be} linked to the \edt{greater} tail's acceleration, and a short rebound at the end of the burst. When the coast starts, the side force drops toward zero. 



\begin{figure}[htbp]
\centering
\includegraphics[width=6.5in]{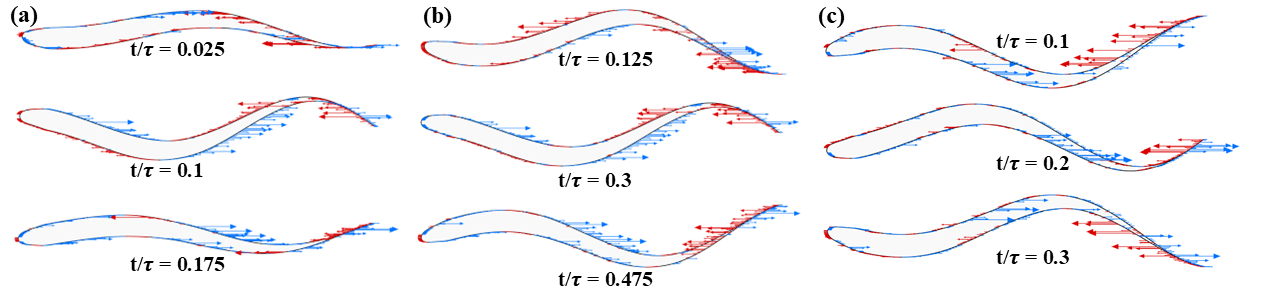}
\caption{\small Horizontal forces applied on the anguilliform swimmer body at different time steps at $\mbox{St} = 0.3$ (a) $\mbox{DC} = 0.2$, (b) $\mbox{DC} = 0.6$ and (c) $\mbox{DC} = 1$ (continuous).}
\label{figure9}
\end{figure}

\begin{figure}[htbp]
\centering
\includegraphics[width=6.5in]{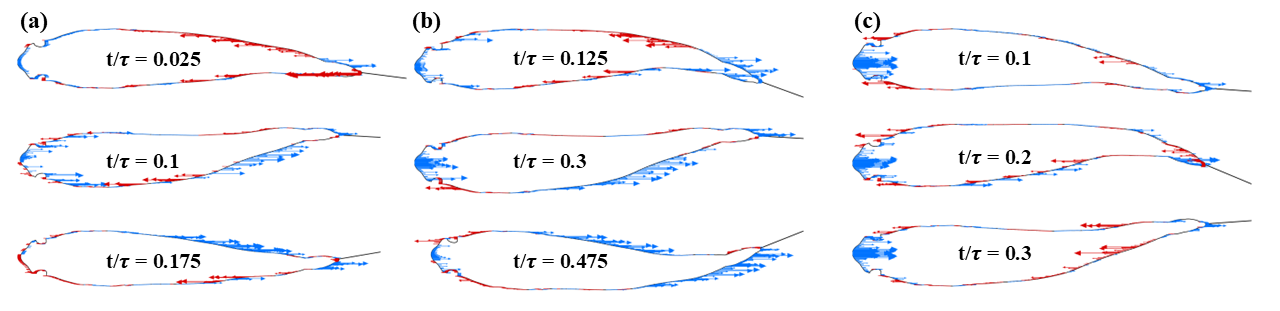}
\caption{\small Horizontal forces \edt{experienced by} the carangiform swimmer body at different time instants for $\mbox{DC} = $ (a) $0.20$, (b) $0.6$ and (c) $1.0$ at $\mbox{St} = 0.30$.}
\label{figure10}
\end{figure}

\edt{From our previously explained observations, an important question arise about why drag is higher at the greater Strouhal number for the intermittent motion of both the swimmers. To understand the physical mechanisms associated with this aspect}, we \edt{now analyze the streamwise forces acting on different part of the swimmers' bodies during their different kinematic maneuvers. For this purpose,} we project the surface pressure onto the streamwise direction, and plot the local horizontal force as short arrows on the body, \edt{considering the mid-plane section of the body in \edt{Figs.~\ref{figure9} and \ref{figure10}}. It is important to mention that we disregard the influence of shear forces in these computations, following the approach of Gemmel et al. \cite{gemmell2016bending} and Du et al. \cite{gemmell2016bending}. Here, the} blue arrows point downstream and mark drag, \edt{whereas the} red arrows point upstream and mark thrust. We show three instants within one cycle \edt{to analyze how the pressure-related forces are exerted on different parts of the body. For} the anguilliform swimmer, \edt{Fig.~\ref{figure9} presents the force profiles for} $\mbox{DC} = 0.2$ \edt{and $0.6$, as well as} continuous undulation at $\mbox{St} = 0.30$. At $\mbox{DC} = 0.2$\edt{,} the burst starts with strong upstream motion near the tail and lower pressure on the \edt{anterior} side \edt{with} the tail and the \edt{posterior} body \edt{generating} most of the thrust \edt{(see Fig.~\ref{figure9}a). Immediately after it happens}, we observe a large \edt{blue arrows} near the tail as the tail slows and the local acceleration reverses. \edt{This} band marks the impulsive drag that raises the drag in burst and coast motion in Fig.~\ref{figure7}$a_1$. As we raise $\mbox{DC}$ to $0.6$, the band of red arrows at the tail remains there but the blue drag band weakens and shortens in time, as shown in the middle row of plots in Fig.~\ref{figure9}. The flow around the mid-body also settles and the patches holding blue arrows shrink. At $\mbox{DC} = 1$ the overall pattern of {blue and red arrows seem} balanced \edt{with} the tail \edt{still producing more} thrust \edt{during the kinematic cycle. Nevertheless,} the amplitudes of the \edt{streamwise pressure-related forces} stay small, and we do not see the \edt{indications for} large impulsive drag events, \edt{previously seen for} low $\mbox{DC}$. 



\edt{For} the carangiform swimmer, \edt{plots in} Fig.~\ref{figure10} show the {instantaneous pressure-related forces on the swimmer's body for} $\mbox{DC} = 0.2$, $0.6$, and $1.0$ at $\mbox{St} = 0.3$. At $\mbox{DC} = 0.2$, the tail produces a strong red thrust\edt{-indicating} band at the start of burst, then a wide blue band over the caudal fin during \edt{the} stroke reversal. That blue-arrow band marks the main drag source in burst and coast for the Jack Fish. As we \edt{increase} $\mbox{DC}$ to $0.6$, the \edt{cluster of blue arrows} weakens and \edt{gets shortened}, and the mid-body carries a larger share of nearly symmetric loads. In the \edt{case for} continuous motion, \edt{the region near the peduncle and the tail experience larger streamwise forces. Here, the flow looks more streamlined, traversing along the body, in comparison to} the eel, because the tail is \edt{able to produce a more axially-oriented wake from the beginning of the motion}. These \edt{physical mechanisms} \edt{explain} the \edt{smaller} side-force and drag slopes \edt{for the Jack Fish with an increasing} $\mbox{DC}$ in \edt{Figs.}~\ref{figure7}$b_1$ and \ref{figure8}$b_1$.


\begin{figure}[htbp]
\centering
\includegraphics[width=6.5in]{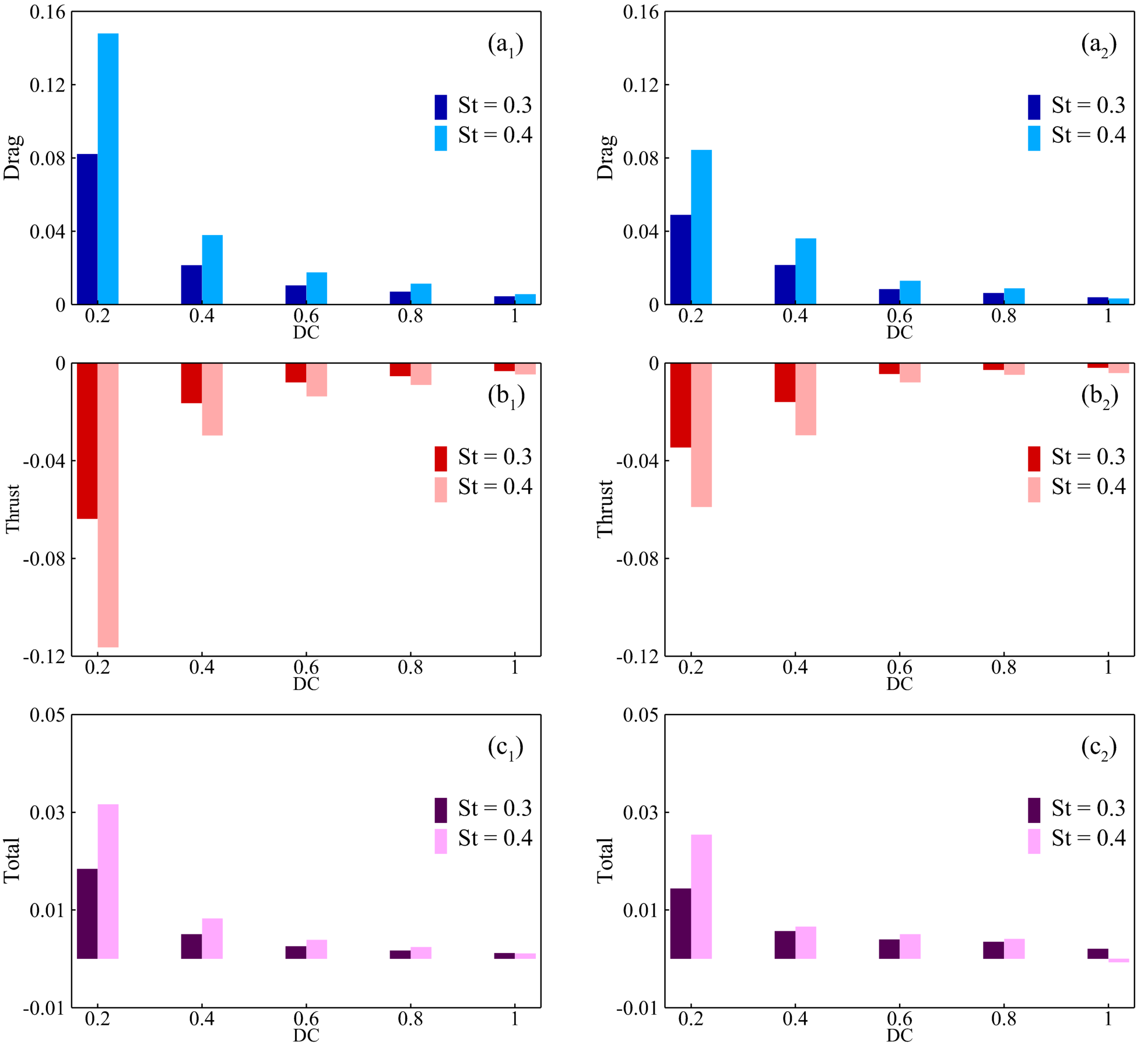}
\caption{\small Cycle-averaged pressure forces versus duty cycle DC for anguilliform ($a_1$,$b_1$,$c_1$) and carangiform ($a_2$,$b_2$,$c_2$) swimmers: drag (top), thrust (middle), and net (bottom) at $\mbox{St} = 0.3$ and $\mbox{St} = 0.4$.}
\label{figure11}
\end{figure}

\edt{Although the overall streamwise force coefficient $C_D$ is already quantified and presented previously, it is insightful to focus solely on quantifying the pressure-component of the forces indicated in Figs.~\ref{figure10} and \ref{figure10}. Therefore,} we plot \edt{the} cycle-averaged drag and thrust caused by pressure along the whole body, and their net \edt{metric} for each $\mbox{DC}$ and $\mbox{St}$ in for both swimmers in Fig.~\ref{figure11}. \edt{In the case of} the anguilliform swimmer (see Figs. \ref{figure11}$a_1$,\ref{figure11}$b_1$, and \ref{figure11}$c_1$), \edt{there are a few clear observations. Looking at the drag profiles (Fig.\ref{figure11}$a_1$ for eel and Fig.~\ref{figure11}$a_2$ for Jack Fish), it is evident that the cases with higher $\mbox{St}$ produces substantially more drag for all $\mbox{DC} < 1$. For the case of $\mbox{DC}=1$, the eel produces slightly greater drag at $\mbox{St}=0.40$ compared to that on $\mbox{St}=0.3$ from the pressure component of the streamwise force. However, the Jack Fish produces significantly less drag at the greater value of $\mbox{St}$. In the thrust profiles (see Figs.~\ref{figure11}$b_1$ and \ref{figure11}$b_2$), we see similar observations for the two swimmers. The only difference is that the Jack Fish produces comparatively larger thrust for $\mbox{St}=0.40$ and $\mbox{DC}=1.0$. The total (net) force in Figs.~\ref{figure11}$c_1$ and \ref{figure11}$c_2$. shows how the overall drag is more for the higher Strouhal number for all $\mbox{DC}<1$ for both anguilliform and carangiform swimmers. These observations also relate very well the profiles of $\bar{C_D}$ presented in Fig.~\ref{figure5}.}


\begin{figure}[htbp]
\centering
\includegraphics[width=6.5in]{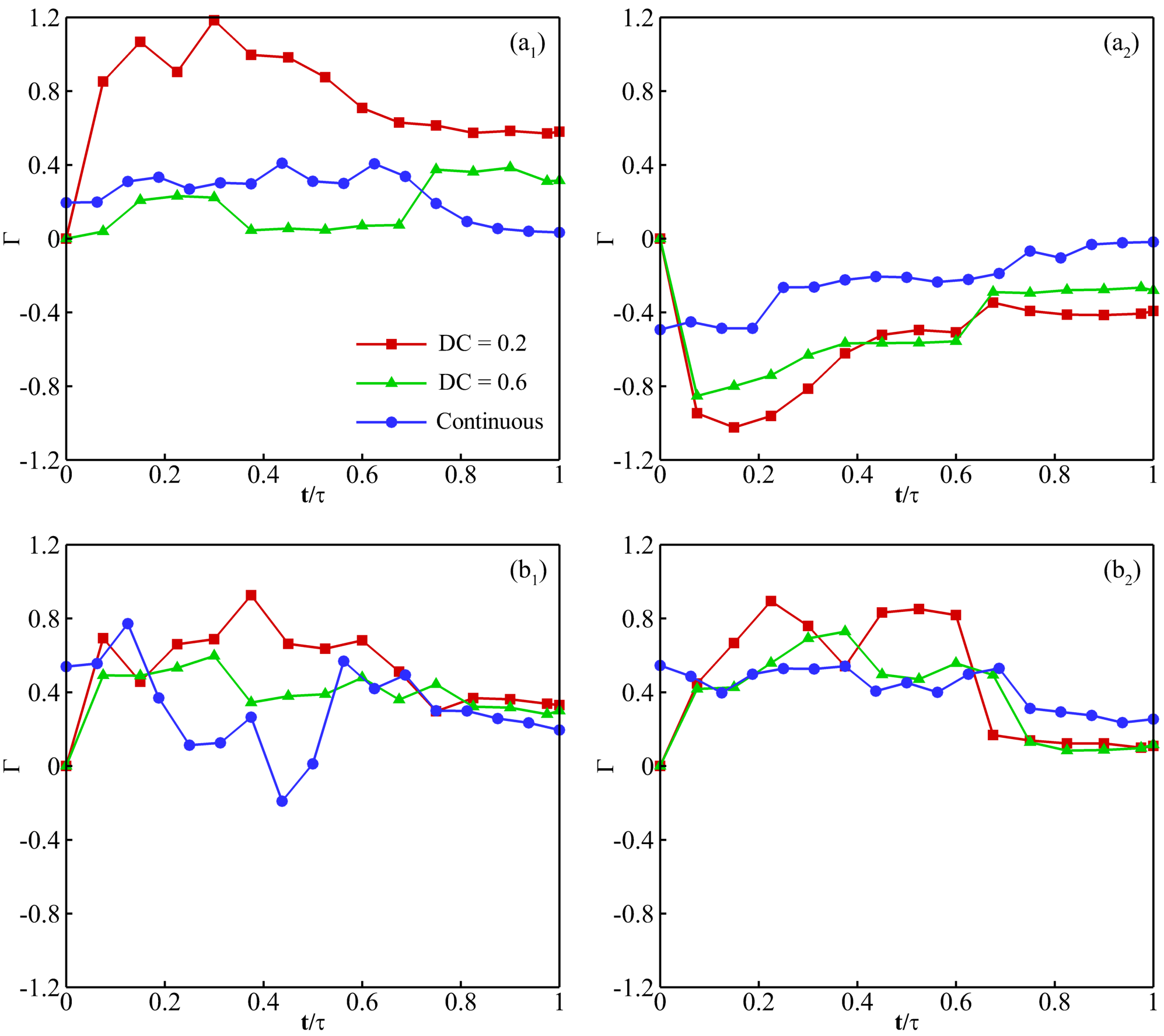}
\caption{\small \edt{Circulation of the ($a_1$) right-sided vortex and ($b_1$) left-sided vortex around the eel, and that of the ($a_2$) right-sided vortex and ($b_2$) left-sided vortex around the Jack fish}}
\label{figure12}
\end{figure}

\edt{The next important aspect of the instantaneous hydrodynamic streamwise force in Fig.~\ref{figure7} relates to larger peaks for low values of $\mbox{DC}$ compared to its larger values. In order to elucidate this phenomenon, it is logical to quantify the strength of the vortices shed from the body under different kinematic conditions, as this metric directly relates to the magnitude of the hydrodynamic forces, experienced by the swimmer. Hence,} we track circulation ($\Gamma$) of the right and left vortices marked \edt{earlier in Fig.~\ref{figure3}} and compare three cases for each gait: $\mbox{DC} = 0.2$, $\mbox{DC} = 0.6$, and continuous \edt{undulation} at $\mbox{St} = 0.3$ \edt{in Fig.~\ref{figure12}}. Please note that Figs.~\ref{figure12}$a_1$ \ref{figure12}$b_1$ provides the plots for instantaneous circulation of the vortices initially shed from the right and left sides, respectively, of the eel, whereas Figs.~\ref{figure12}$a_2$ \ref{figure12}$b_2$ provide $\Gamma$ for their counterparts around the Jack Fish. It is also important to mention that the purpose of these plots are to quantify the strength of the first vortices shed from the two sides of the swimmers, regardless of their orientation. All these plots make it evident from the magnitudes of $\Gamma$ that the vortices formed and shed on the two sides of both the swimmers are stronger for small $\mbox{DC}$. They also keep their comparative strength intact while traversing in the wake after getting shed from the bodies and interacting with other secondary flow structures in the wakes.

\section{Conclusions}

Undulatory \edt{marine} swimmers attain \edt{their hydrodynamic} performance by \edt{controlling} how near-body vortices grow, \edt{get} shed, and \edt{form} the wake. \edt{In this study, we examine how duty cycle, as an important feature of the burst-and-coast motion of carangiform and anguilliform swimmers, influence vortex dynamics and formation of the wakes and affect the generation of unsteady hydrodynamic forces}. This work reveals the formation of a bow-shaped wake for smaller values of $\mbox{DC}$, which further becomes more streamwise as the two rows of vortices from the left and right sides of the swimmer approach each other with an increasing value of $\mbox{DC}$. Our results provide evidence for how $\mbox{DC}$ controls the bow angle in such cases. \edt{Secondly, we explain the greater drag in burst-and-coast motion for the higher Strouhal number, that is contrary to what is usually observed for continuously undulating swimmers}. \edt{Furthermore,} the eel forms a largely spread wake with higher bow angles during the intermittent swimming, where the carangiform swimmer demonstrates lesser sensitivity to this feature as a function of $\mbox{DC}$ primarily due to the substantial motion of the posterior parts of the body. We also explain the mean and instantaneous hydrodynamic forces for different values of duty cycle and Strouhal number, that provide new insights about the hydrodynamic behavior of two entirely different kinematics and morphologies of biological swimmers. 




\section*{Acknowledgment}

MSU Khalid acknowledges funding support from the Natural Sciences and Engineering Research Council of Canada (NSERC) through the Discovery and Alliance International grant programs for this work. A. Tarokh also thanks NSERC for their support though the Discovery grant. The simulations reported in this work were performed on the supercomputing clusters administered and managed by the Digital Research Alliance of Canada. 

\vskip6pt

\bibliography{references}

    \end{document}